# USING CONTEXT TO IMPROVE THE EVALUATION OF INFORMATION RETRIEVAL SYSTEMS

Abdelkrim Bouramoul[1], Mohamed-Khireddine Kholladi[1] and Bich-Lien Doan[2]

[1] Computer Science Department, Misc Laboratory, University of Mentouri Constantine. B.P. 325, Constantine 25017, Algeria
`a.bouramoul@yahoo.fr` , `kholladi@yahoo.fr`.

[2] Computer Science Department, SUPELEC. Rue Joliot-Curie, 91192 Gif Sur Yvette, France.
`bich-lien.doan@supelec.fr`

*ABSTRACT*

*The crucial role of the evaluation in the development of the information retrieval tools is useful evidence to improve the performance of these tools and the quality of results that they return. However, the classic evaluation approaches have limitations and shortcomings especially regarding to the user consideration, the measure of the adequacy between the query and the returned documents and the consideration of characteristics, specifications and behaviors of the search tool. Therefore, we believe that the exploitation of contextual elements could be a very good way to evaluate the search tools. So, this paper presents a new approach that takes into account the context during the evaluation process at three complementary levels. The experiments gives at the end of this article has shown the applicability of the proposed approach to real research tools. The tests were performed with the most popular searching engine (i.e. Google, Bing and Yahoo) selected in particular for their high selectivity. The obtained results revealed that the ability of these engines to rejecting dead links, redundant results and parasites pages depends strongly to how queries are formulated, and to the political of sites offering this information to present their content. The relevance evaluation of results provided by these engines, using the user's judgments, then using an automatic manner to take into account the query context has also shown a general decline in the perceived relevance according to the number of the considered results.*

*KEYWORDS*

*Contextual Evaluation, Evaluation Campaigns, Relevance Judgments, Information Retrieval, Web Search Engine.*

## 1. INTRODUCTION

Information Retrieval (IR) is now an activity of great importance to the extent that it became one of the most important actors in the rapid development of new information and communication technologies. It must be possible, among the large volume of documents available, finding those that best fit our needs in the shortest time, for this purpose, information retrieval tools have been developed to help locate information in a closed corpus of documents or among the entire document available on the web. Consequently several questions arise about these information retrieval tools, particularly in terms of their performance and the relevance of the results that they offer.

It is therefore in the field of evaluation of information retrieval systems and more specifically that of the contextual evaluation that our work falls. After a deep investigation around research and synthesis activities we realized that despite the abundant literature produced in this area dealing with both experimental results and methods that provide evaluation criteria and metrics of relevance, few of these methods are interested in the consideration of the context during the evaluation process. Our contribution is guided by two main reasons; firstly the lack that we observed around the context-based methodologies for measuring the quality of information retrieval tools, this finding is reinforced by the work of [1] and [2]. And secondly by the requirement to which we are confronted recently after conducting work in the field of the consideration of context in information retrieval systems [3]; Where we have failed to find a contextual evaluation protocol to validate of our proposal. This work will therefore be a logical continuation of what has been done before, and a promising way to cover the process of the contextual evaluation of Information Retrieval Systems (IRS).

This paper is organized as follows; we start first by giving a definition of the concept of context and its use in the field of information retrieval, we then present an overview of classical approaches for evaluating information retrieval systems and we focus on the limits and shortcomings faced by these approaches. In the next section we discuss our contribution by giving an overview of the contextual evaluation approach that we propose and describing its principle and its techniques. To demonstrate the applicability of the proposed approach we present the experimentation that we conducted to evaluate the performance of the three search engines, Google, Yahoo and Bing. We finally discuss the results and we end with a conclusion.

## 2. HOW CONTEXT CAN BE USED IN IR

### 2.1. Definition of context

The context is not a new notion in computer science: from the sixties, operating systems, language theory and artificial intelligence already exploited this concept. With the emergence of information retrieval systems, the term was rediscovered and placed at the core of the debates without making subject of a consensus, clear and definitive definition. However, analysis of existing definitions in the literature leads to two conclusions:

- "*There is no context without context*" [4]. In other words, the context does not exist as such. It is defined or it emerges for a purpose or precise utility.

- "*The context is a set of information. This set is structured, it is shared, it evolves and serves the interpretation*" [5]. The nature of information and interpretations got from it depend on the purpose.

In information retrieval, the context is defined as "All cognitive and social factors as well as the user's aims and intentions during a search session", [6]. Generally speaking, the context includes elements of various natures that delimit the understanding, the application fields or the possible choice. The most commonly cited elements concern the spatiotemporal data (location, time, date) or specific knowledge in relation to the studied area. But rarely, we see the use of elements concerning the emotions, state of mind cultural information [4].

### 2.2. Use of context in information retrieval

In information retrieval, context can be used at three different stages depending on the progress of the research process. The context may be considered before the research process, during the research process, or at the end of the research process:

#### 2.2.1. At the beginning of the search process

The context can be used in a pre-research phase to solve the problem of ambiguous terms in the query and improve the quality of results returned by the system. We can for example assist the user in formulating his query by asking him to clarify, according to the context of the current search session, the sense of an ambiguous word using a thesaurus or ontology, We quote in this category the work of [7] that uses an ontology with equivalence and subsumption relationships for extracting terms to be added to the initial query.

Another simpler way to use the context in a pre-search phase is to use it in the introduction of booleans constraints on the existing algorithms of information retrieval, these algorithms can also consider the spatiotemporal context in which the continuous values can be described in a non-specific manner to different granularity levels [8]. For example, an event can take place at 9:57, at about 10am or in the morning. In this case, the context can be used for selecting the appropriate representation.

### 2.2.2. During the search process

The context can also be considered in the interactions with the system. Indeed, in an information retrieval process, is the interaction that makes possible the real exploitation of the displayed results. The user is particularly adept to extract information from an environment that control directly and actively compared to an environment that he can only observed it passively [9]. The context at this level depends on the user action in a given situation, on the feedback, on the relevance judgments that are related to characteristics of different users' situations, to the multidimensional research strategies, and other informational practices in information retrieval.

### 2.2.3. At the end of the research process

The context may finally be considered in a post-search phase after obtaining the results by using the relevance feedback principle. The idea of this technique is to achieve a first search using only the query terms, and the user can then indicate which are, among the best document of this first search, those that are relevant and those which are not, and the system uses this information to refine the search by changing the weights of query terms using an automatic learning methods as in the work of [10]. Another way to use the context with the relevant feedback has been proposed more recently in our work [3] where we propose a contextual query reformulation based on user profiles using the concept of static and dynamic context to minimize the user intervention in the process of reformulation.

## 3. RELATED WORK

The classic evaluation of information retrieval systems is based on the performance of the systems in themselves; it is quantitative and is based on work done in the sixties at Cranfield (United Kingdom) on indexing systems [2] . This type of approach provides a comparative evaluation basis of the effectiveness of different algorithms, of techniques and/or of systems through common resources: test collections containing documents, previously prepared queries and associated relevance judgments, and finally evaluation metrics essentially based on the recall and precision. [11]

### 3.1. Evaluation campaigns

The evaluation campaign represents the current dominant model. Indeed, it is on the experience of the Cranfield tests that was based the NIST (National Institute of Science and Technology) to create the TREC evaluation campaign (Text REtrieval Conference) in 1992. The TREC campaigns have become the reference in the evaluation of systems but we can also quote the CLEF Campaigns (Cross-Language Evaluation Forum) which specifically relate to the multilingual systems, the NTCIR campaigns on the Asian languages, and Amaryllis, specializing in French systems.

### 3.1.1. The TREC evaluation campaign

This is a series of annual evaluation of information retrieval technologies. The TREC is an international project initiated in the early 90s by the NIST (Institute in the United States), in order to propose homogeneous means for the evaluation of documentation systems on a consistent basis of documents. The participants are usually researchers for large companies which offer systems and that want to improve it, small vendors that specialize in the information retrieval or academic research groups.

The TREC is now considered as the most important development in experimental information retrieval. The TREC program has had a very important impact in the field, and remains the most cited and used by the information retrieval community. The main explored tracks are filtering, research (or ad hoc task), interactive, Web and question-answering. For 2010 TREC has focused on the following tracks: The blog, chemical IR, entity, legal, relevance feedback, and session tracks.[1]

### 3.1.2. The CLEF campaign

In 2000 is launched the European project of evaluating information retrieval systems, this project is called CLEF (Cross Language Evaluation Forum). The objective of the CLEF project is to promote research in the field of multilingual system development. This is done through the organization of annual evaluation campaigns in which a series of tracks designed to test different aspects of mono- and cross-language information retrieval are offered. The intention is to encourage experimentation with all kinds of multilingual information access – from the development of systems for monolingual retrieval operating on many languages to the implementation of complete multilingual multimedia search services. This has been achieved by offering an increasingly complex and varied set of evaluation tasks over the years. The aim is not only to meet but also to anticipate the emerging needs of the R&D community and to encourage the development of next generation multilingual IR systems.

CLEF 2009 offered eight main tracks designed to evaluate the performance of systems, the most important of these tasks are: Multilingual textual document retrieval, interactive cross-language retrieval, cross-language retrieval in image collections, intellectual property and log file analysis [12].

## 3.2. Limits of classic approaches for evaluating IRS

Despite the popularity and recognition of these two evaluation campaigns that are TREC and CLEF. These approaches for evaluating information retrieval systems have some limits particularly with regard to the user consideration, the constitution of the queries corpus but also about the evaluation itself.

To better identify the limits of classic approaches for evaluating information retrieval systems, and basing on the work of [1] , [2], [13]. A Synthesis of this work has allowed us to define three classes of problems. Each class is related to an actor who is generally present around an evaluation process; these limits are those related to the absence of the user in the evaluation process, those related to the relevance judgments, and finally the limits related to the corpus of documents and queries.

### 3.2.1. Limits in relation to the user

We can reproach these evaluation approaches to be artificial and arbitrary. While TREC has effectively improved the efficiency of the system, the notion of the end user implies personal knowledge, experience and different research capabilities, for which the system evaluation does not care. Indeed, such evaluations ignore the context in which the research is conducted since they are not performed in real use situations. In this context [13] asserts that the absence of the user in the evaluation process is one of the first and probably most important critique of classic approaches using criteria other than those of the recall and precision when they initiate or end a search session.

---

[1] TREC web site : http://trec.nist.gov/

### 3.2.2. Limits in relation to judgments of relevance

The relevance is a subjective notion and it seems unthinkable to measure it without being arbitrary. We also note that the relevance judgments in TREC operate on a binary manner: a document is considered as relevant or irrelevant. Yet this is obviously not always the case, some documents are more relevant than others who are also relevant. These degrees of relevance are still dependent on the mindset of the person who actually needs these documents. This finding is validated by the work of [1] showing that the relevance considered in the classic evaluation of IRS is thematic, independent of context, of the research situation and interests of users. Similarly the work of [13] have shown that relevance judgments should be revised in the sense that they are stable and do not vary over time, and that they are assigned independently of each other.

### 3.2.3. Limits in relation to the corpus of documents and of queries

In the traditional corpus, a document is a text in itself, and the evaluation is made compared to the number of documents found, but in general, a user is not looking documents but information, and documents never contain the same amount of information. Similarly for corpus of queries where the query is a need for information expressed in natural language. However, the representation of the information needs of the user is itself a problem. The IR task becomes a task of know ask questions to these systems because the differences are significant between what we think and what is interpreted. [1] Notes that in the batch mode of evaluation protocols, queries are assumed to represent alone the user. Consequently the direct users having made these queries, their interests and their interactions with the IRS does not form part of the collection.

This critical finding prompted our reflections around an appropriate approach for contextual evaluating of the information retrieval systems. In the rest of this paper, we describe our approach for taking context into consideration during the evaluation process.

## 4. DETAILED PRESENTATION OF THE PROPOSED APPROACH

Our evaluation approach is to evaluate the performance of the tool used for information retrieval and measure the quality of services that it offers in one side, and on another side to evaluate the relevance of the results that it returns. It takes the user into consideration during evaluation in the sense that it contributes to the evaluation process by giving his relevance judgment according to his information need. The proposed approach therefore consists of three parts: evaluation of performance of the search tool, evaluation of the relevance of results compared to the query, and finally evaluation of the relevance by the user's judgments. Fig. 1, summarizes the three levels of evaluation and illustrates the link between the context type and the evaluation level.

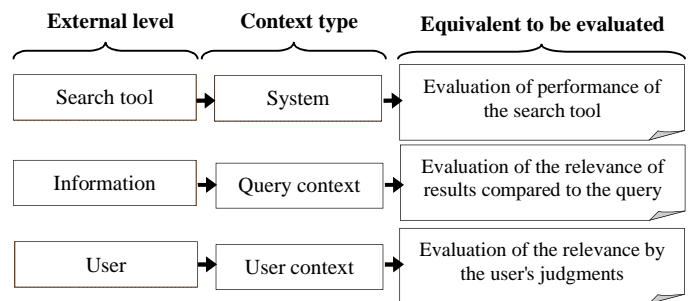

Figure 1. Link between the context type and the evaluation level

We chose to consider three types of context modelled in our approach with three complementary evaluation levels, these are:

- *System context:* at this level it comes to diagnosing performance, characteristics, specifications and behavior of the search tool for the considered query.

- *Query context:* this is to measure in incremental way at what point the returned results reflect the user's information need.

- *User context:* in addition to the score given by the system, it comes to responding to the question, how the user appreciates results. This information is subsequently capitalized as historic for reuse it in a future evaluation sessions.

## 4.1. Evaluation of performance of the search tool

This is the first component of our approach; the evaluation referred to this level is based on a number of criteria summarizing the problems generally encountered by users during a search session. The criteria that we have defined depend on the nature of the manipulated information, of the source of this information, and finally of the mechanism used to retrieve this information. The values assigned to these criteria are automatically calculated by the system soon obtaining results provided by the search tool. The estimation of these values gives subsequently an overview of the quality of the search tool independently of the relevance of the results that it returns. These criteria are the following:

*The redundant results:* This involves measuring the ability of the search tool to discard the redundant results. This means that the search tool should return only once the results coming from the same site but with different pages.

*The dead Links:* A dead link is a link that leads to a page that does not exist, that it has been moved or deleted. In general, the browser returns in this case the error codes '404'. Evaluate this criterion consists to underscore the ability of the search tool to detect them.

*The parasites pages:* They include advertising pages and pages that can identify, for example, only promotional links. These pages provide no useful information to the user and generally make false results. Their elimination depends to the performance of the 'crawler' search tools, and hence the quality of the algorithms used by each search engine.

*Response time:* This is the time consumed by the search engine to return the query's results; it is one of the most important aspects. More response time is short, better are performance of search tool.

## 4.2. Evaluation of the relevance compared to the query

This is the second part of our contextual evaluation approach; this is the weighting, by increasing the number of terms, of the query words compared to the words of the returned documents. This includes choosing the weighted terms in the first time, then apply the formula that we propose an incremental way versus the number of words forming the query.

### 4.2.1. Weighted terms choice, an incremental weighting

In a process of information retrieval queries are created by the user, it reflects an information need, and they are composed of one or more words depending to the necessity to satisfy the deficiency noted in information, a lacuna or a defect. The groups of words in a query are often more semantically rich than the words that compose it taken separately, and can therefore better respond to what users expect.

In our approach, we have chosen to define several hierarchal levels during weighting according to the number of words forming the query. Each level is composed of one or more words (a group of words) starting from the query formulated by the user. The incremental weighting by increase of query terms instead of a classic weighting of each word separately allows better take into consideration the query context during the evaluation. For example, assuming that the query sent by the user is 'contextual evaluation of information retrieval systems', documents containing the group of words : 'contextual evaluation of information' or 'contextual evaluation' are certainly nearest to what the user expect compared to those in which we find the words: 'contextual', 'evaluation', 'information', 'retrieval' or 'systems' taken separately.

### 4.2.2. Relevance Calculating, a contextual formula

Once the groups of words to be weighted are defined, it comes to assigning a weight that determines their importance in the document. We have therefore developed a weighting formula that takes into account the context of the query in terms of number of words composing it. This formula is inspired from the TF IDF weighting **[14]** to which we added two dimensions; the document length and the hierarchy of words groups according to the length of the query. So, it is incremental and is defined as follows:

$$W(R,D) = \sum_{R' \in R} \left[\frac{W(R',D)}{length(D)}\right] * \left[\frac{length(R')^2}{length(R)}\right] * \log_2\left[\frac{TNRD}{NDWDR'}\right]$$

With:
- R: The set of query terms.
- R': The terms of the words group to weighted.
- W (R', D): The frequency of R' in the document D.
- Length (R): the query length.
- Length (R'): the length of the words group to weighted.
- Length (D): Length of the document.
- TNRD: Total number of returned documents.
- NDWGR': Number of documents containing R'.

## 4.3. Evaluation of the relevance by the user's judgments

When an IRS returns a document to the user, this one recovers information. This information is important for a given user; it is possible that the same information can make a greater or lesser importance, generate a more or less bright interest depending on the individual and the context of use. The information has therefore importance for a given user in a given context and is the user that determines the actual adequacy of results returned by the search tool with its information need. Based on this principle and to allow consideration of the user's judgments during the evaluation, we use an adaptation of our approach proposed in [3] which is to model the user by a static and dynamic context. The migration of our approach from taking into account the context in information retrieval to its consideration in the evaluation process requires a redefinition of the concept of static and dynamic context to make them usable for evaluation.

### 4.3.1. Static context

These are the personal characteristics of the user that can influence the research context. This information is stored in the user's context base during the first connection to the system. For this purpose we have identified four categories of information relating to the static context, this information is summarized in:

- Connection parameters: e-mail and password.
- Personal characteristics: name, country, language,...
- Interests and preferences: domains, specialty,...
- Competence expertise level: profession, level of study,...

After having recovered the static context, the user can formulate his query and the search tool takes charge to returned suitable results.

### 4.3.2. Dynamic context

In order to optimize the reuse of the user's judgments and facilitate their understanding, this second component of context aims to associate the relevance judgments with the user's context. The principle is as follows; at the end of each search session the recovery of the dynamic context is performed and this by allowing users to express their judgments of relevance regarding to the documents returned by the search tool. This judgment by the user is to vote on a scale from 0 to 5, where 0 corresponds to a document completely useless or off-topic, 5 corresponding to a document that responds perfectly to the asked query. The evaluation is activated automatically whenever the user expresses a judgment. Finally and based on relevance judgments assigned by the user, the system recalculates the relevance value of a result and the evaluation of the search tool is carried out by updating the basis of the user's contexts.

## 5. APPLICATION OF THE PROPOSED APPROACH TO THE EVALUATION OF SEARCH ENGINES

To prove the applicability of the proposed approach, we used it to for the contextual evaluation of search engines. Our choice was set on three search engines (Google, Yahoo and Bing). This choice is motivated by their popularity in the web community on the one hand, and the effectiveness of their research nuclei and the degree of coverage that they provide in response to a request on the other hand.

We therefore propose to set up a system conducting an open search on the web, and perform by following the evaluation of the results returned by each search engine. To this end we use the three levels of the contextual evaluation approach that we have proposed. This system should allow:

1) Make the same set of queries to the three search engines Google, Yahoo and Bing.
2) Retrieve the results returned by each search engine;
3) Check the informational content of all the resulting pages;
4) Capture the user's static and dynamic context for the current search session, and used it for the evaluation of the results by the user's judgment;
5) Measuring the degree of relevance of results returned by each engine taking into account the context of the query by the incremental application of the proposed formula.
6) Diagnose performance, characteristics, specifications and behavior of each search engine taking into account its context accordance with what has been proposed in the third level of our approach.
7) Coupling of the relevance scores obtained in the three evaluation levels for each search engine and thus obtained the final evaluation.

The system consists of two main modules: a first module for managing interactions between the user and the search engine (identification and search), and a second which covers the three levels of evaluation described in our proposal. These two modules are closely interrelated in the sense that the outputs of a module are the inputs of the other. We present in the in what follows modules components the system and we illustrate the functionalities offered by each of them.

### 5.1. Managing of users / search engine interactions module

We are interested to evaluating of the quality of search engines and the relevance of the results that it returns. A preliminary phase to this evaluation is absolutely necessary, it involves taking into account the user's information need in the form of a query and then interrogate the search engine selected to retrieve results to be evaluate. The managing of users/search engine interactions module supports all interactions between the user and the search engine for the connection to the system until the results deliverance.

It takes care capturing of the user's static context, managing of its identification, He also manages the transmission of the user request to the search engine and retrieval of results, and finally, it communicates these results to the evaluation module. This module consists of two complementary processes:

### 5.1.1. The static context capturing process

The static context previously defined during the presentation of our approach is represented by the user profile. The latter is the source of knowledge defining all users' aspects and which can be useful for system behavior. The user profile data comprising the static context can be indicated by the user himself, learned by the system during use or indicated by selecting an existing profile created by experts. In our case, we construct the static context of the user at the first connection to the system. This construction is done by asking the user to fill the four categories of information defined previously.

The categorization of users has the advantage of having typical information with the opportunity to refine it as and when. Once the identification made the user can conduct open research on the web.

### 5.1.2. The search process

We opted for a system that offers an open search on the web using the following principle; after connecting to the system, the user expresses his information need as a query. The research process therefore takes as input the query and gives to the user the ability to choose one of three search engines that the system proposes (Google, Yahoo, and Bing), the search operation is initiated by running in parallel the nucleus of each search engine with as only parameter the user query. The obtained result is finally communicated to the user and the evaluation module. This process also calculates the response time of each search engine. Figure. 2, shows module for managing interactions between the user and the search engine and illustrates the operating principle of its two processes.

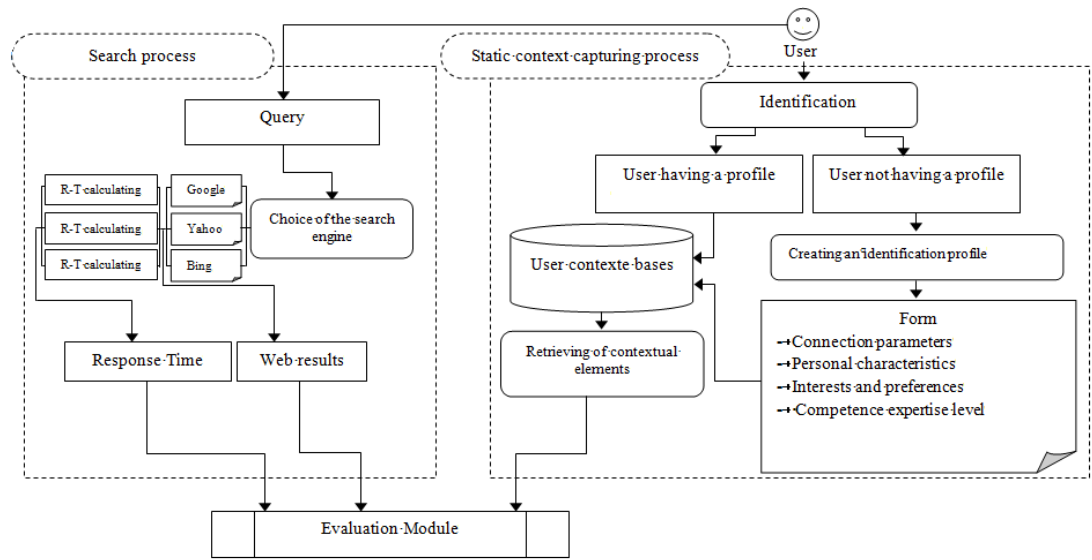

Figure 2. Managing of users / search engine interactions module.

## 5.2. Evaluation module

To precede with the evaluation of the three search engines, the system retrieves the results of each of them and performs their analysis. The contextual evaluation module consists of three processes representing the three levels of evaluation of the approach that we propose. These processes are respectively; a first for the performance evaluation of the search engine, a second process for the automatic evaluation of the relevance of results returned by this engine and finally a process for the evaluation of the relevance by the user's judgments. Figure 3, summarizes the evaluation approach applied to search engines.

### 5.2.1. The performance evaluation of the search engine process

This process diagnostic performances and characteristics of each search engine based on the criteria developed in our approach. It takes place according to the following steps:

*Extraction of the link list:* As soon as the search engine displays the results in response to a user query, the system automatically retrieves the list of links 'url' related to each result and performs the appropriate treatment according to the page content.

*Detection and calculates of the redundant links:* it concerns analysis of the links list to detect those that are redundant and calculate their number. If there is no redundant link the note will be equal to the number of analyzed links, Otherwise the note decrease by one as much as there are redundant links.

31

*Detection and calculates of the dead links:* The detection of such links is by opening a connection with all the links recovered, if the open operation fails, the link is considered as dead. For assigning the final note, the principle is the following; calculate the number of dead links and assigns the note of '0' if links are dead and the note of '1' otherwise.

*Detection of parasite pages:* it includes pages that do not contain at least one of the query terms in the returned results. The detection operation is to calculate the number of occurrences of each query word in the documents. If the frequency of each word is equal to '0' then the result is considered as a parasite page.

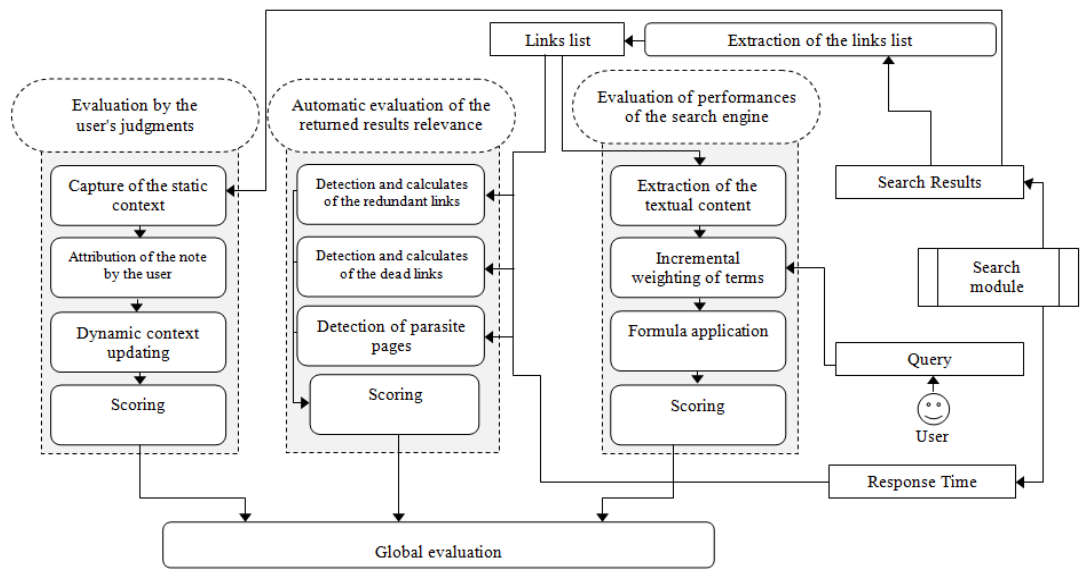

Figure 3. Summary of the evaluation approach applied to search engines

**5.2.2. The automatic evaluation of the returned results relevance process**

This process is interested to the automatic measuring of how the results delivered by the search engine match the user's information need. Consistent with the approach that we propose it unfolds according to the following steps:

*Extraction of the textual content:* This operation is carried out from the previously retrieved links list. The idea is to open the web page corresponding to each 'url', and retrieve its textual contents using a parser developed for this purpose. We have implemented a parser for each search engine because the html tags are different from one engine to another. The extracted content is sent for analysis.

*Incremental weighting of terms:* Once the textual content is retrieved, it comes to calculate the occurrence frequency of the query terms in the different returned documents. The occurrences calculation for each hierarchy level of the considered query is respected according to the number of words forming the query. In other words; the 'n' words composing the query are regarded as one term in the first time and its frequency in each result is calculated, then the 'n-1' query words become the considered term and the frequency is also calculated, and the operation continues until there remains only one word, its frequency is calculated and the incremental weighting comes to an end. In the case where the user wants to search with the exact expression the calculates of frequency is carried out once with the entire query as a term.

*The formula application:* this stage of the evaluation process is to apply the formula developed in our proposal. This formula takes as input for each level of hierarchy group of words forming the query: its occurrence frequency, its length, the document length, the query length, the total number of analyzed documents and the number of documents containing that word group. It produces as output a weight representing the relevance of the result according to the query.

**5.2.3. The evaluation of the relevance by the user's judgments process**

This process is to engage users in real search situations. In this context is the user's relevance judgment that determines the performance of the search engine. According to our evaluation approach, each user is characterized by his static context defined at the first login to the system. After login, the system retrieves the judgments made by the user in previous search sessions to update the dynamic context of judgments. The latter is constructed progressively at the end of each search session by allowing the user to give his opinion after having consulted the returned results. Finally and basing on the user's relevance judgments, the process recalculates the results relevance score and the evaluation of the search engine is then updated.

# 6. RESULTS AND DISCUSSION

## 6.1. The used protocol

To measure the contribution of our approach to the search engines evaluation, we use an extension of the evaluation scenario proposed in [15].The evaluation was conducted with the help of 24 students from the second year license STIC (Science and Technology of Information and Communication) at the Mentouri Constantine University, playing the role of users. The goal was not to make an evaluation by experts but by a basic public, reasonably familiar with search engines. 6 topics were chosen, to reflect diverse fields of use. These topics are: News, Animals, Movies, Health, Sports and Travel. Each topic was assigned to a group of 4 students who chose freely 5 queries. For example, for the sports topic, the chosen queries were as follows:

- World Cup 2010.
- France cycling tour
- Formula 1 racing cars
- Famous football players
- Roland-Garros tournament.

The queries were submitted to different engines, and the first two pages containing the 20 results were archived for each query and each search engine. In total, 1800 'url' ware retrieved (6 topics x 5 queries x 20 results x 3 search engine) and organized in the form of triplet (Query, url, page content). Finally the set of triples has been communicated to the system for analysis and evaluation.

## 6.2. Performance of search engines (system context)

We present in Table 1, the obtained scores for the performance evaluation of the three search engines, and Figure 4, gives a graphical interpretation of these results.

Table 1. Search engines performance evaluation

| Search engines | Performance | | | |
| --- | --- | --- | --- | --- |
| | Dead Links | Parasites Pages | Redundant Results | Average Response Time |
| Google | 2,03% | 5,30 % | 4,04% | 0,17 Sec |
| Yahoo | 2,13% | 10,19 % | 4,81% | 0,21 Sec |
| Bing | 1,67% | 8,64 % | 5,32% | 0,22 Sec |

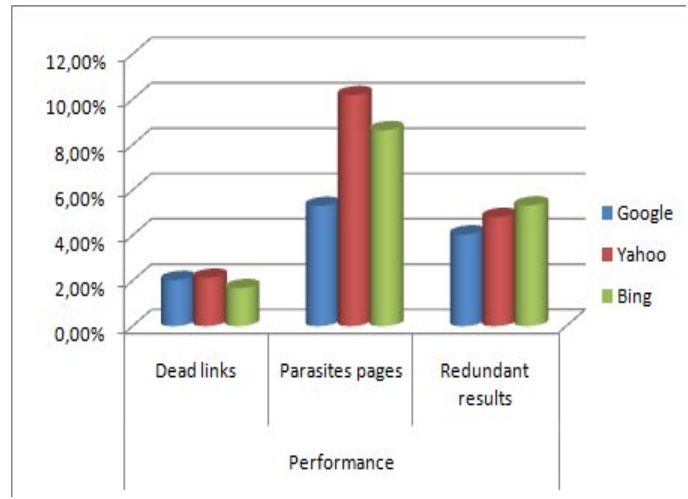

Figure 4. Search engines performance evaluation.

### 6.2.1. Results analysis for dead links

The rate of dead links is low; this is explained partly by the fact that the automatic used procedure tries up to three attempts separated by a delay of few minutes on failure, and secondly by the fact that a number of servers do not return the error code 404 'Page not found' when the page no longer exists, but a normal HTML page with an ad hoc message, which cannot be interpreted as an error only by a human reader. We note also that 71% of dead links returned by Yahoo and 79% of those returned by Google are caused by the Amazon web site which, for unknown reasons, returned an error code during the experiment. Finally, Bing has got the best score with only 1.67% of dead link.

### 6.2.2. Results analysis for the parasites pages

They were considered as parasites the links referring to the commercial sites offering online purchases or transactions. The obtained scores have been variable depending on the search engines, and we notice that they have different strategies to exclude the parasite pages. Among the commercial sites that appear several times we notice two companies: Amazon and E-Bay. Their association with the different engines is interesting to be study. Google and Yahoo are strongly associated with Amazon, while Bing prefers Ebay. Overall, it is Google that returns the fewest links to commercial sites with 5.30%.

### 6.2.3. Results analysis for redundant results

We find that the ability of the three search engines to eliminate the redundant results varied according to the type of queries. The results also showed that the majority of redundant links returned by Google and Yahoo comes from the use of Wikipedia. Of the 20 analyzed results, Google returned 4.04% redundant links whose 80% from Wikipedia, and Yahoo 4.81% redundant links whose 78% from Wikipedia. The results also showed that some web site offer a link type named aliases to avoid redundant links. An alias link type is a copy of main link, with the same URL, but it is not considered by search engines as an attempt to index content abusively.

**6.2.4. Results analysis for the average response time**

This criterion measures the time consumed by the search engine from the query transmission until the results are displayed, it depends heavily on internet connection speed and the power of the machine. To ensure homogeneity when calculating the response time, all queries have been tested on the same machine with the same speed internet connection. The obtained results show that the average response time is almost identical in the three search engines. However, we note that Google top the list with an average speed of 0.17 seconds, this can be explained by the power of the PageRank algorithm used by this engine.

**6.3. Relevance by the user's judgments (user's context)**

We are interested in the relevance judgments given by the user to the first result returned by each search engine (R@01). The latter is of particular importance, since it is the closest link clicked by users. The 24 students also expressed their relevance judgments for 5, 10, 15 and 20 first retrieved documents (R@5, R@10, R@10, R@15, R@20). At each level of relevance, a note of 0-5 was assigned by each student. 0 corresponding to a document completely useless or off-topic, 5 corresponding to a document responding in a perfect way to the question. Table 2, shows the obtained scores.

Table 2. Evaluation of the relevance by the user's judgments

| Relevance level | Search engines | | |
| --- | --- | --- | --- |
| | Google | Yahoo | Bing |
| R@01 | 3,15 | 2,92 | 2,70 |
| R@05 | 2,79 | 2,14 | 2,58 |
| R@10 | 2,34 | 2,51 | 2,16 |
| R@15 | 2,00 | 1,83 | 1,72 |
| R@20 | 1,91 | 1,77 | 1,69 |

The overall scores obtained by each search engine for the 20 results are extremely low, since no motor reaches the average note of 2.5 at R@20. The search engine that had the best note of 1.91 is Google. The situation is remarkably improved if one considers only the first result R@1; the three search engines are exceeding the average.

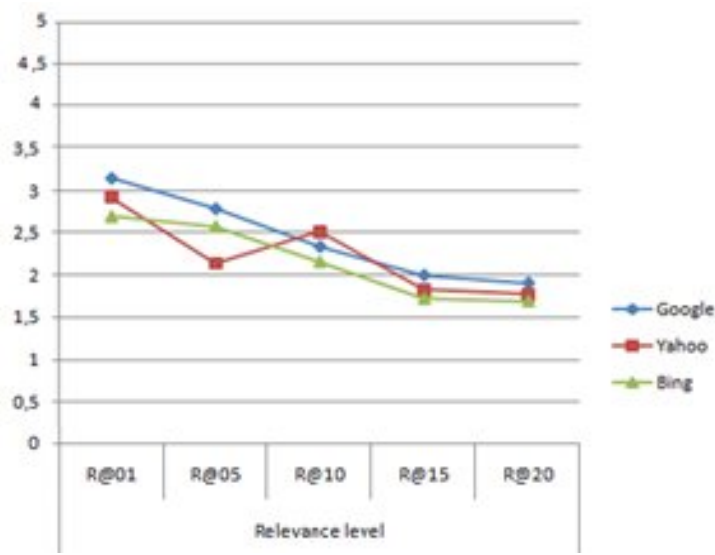

Figure 5. Evaluation of the relevance by the user's judgments

In figure 5, shows the average note according to the level of relevance of results for each search engine. We find a general decline in perceived relevance depending on the number of the considered results, except for Yahoo, which dates back to the average when considering the relevance of the first 10 results R@10, suggesting that the ranking algorithm is not optimal for this engine, or that the result is disturbed by the merging of commercial web sites.

**6.4. Relevance of results according to the query (request context)**

Using our formula, we calculate the relevance of the first 20 returned results according to each of the 30 queries, and that for the three search engines. A note average for each group of 5 queries in the same topic was calculated and the obtained score was rounded to a note on 10. The overall results are summarized in Table 3, and Figure 6, gives a graphical interpretation of these results.

Table 3. Evaluation of the results relevance according to the query

| Queries category | | Search engines | | |
|---|---|---|---|---|
| | | Google | Yahoo | Bing |
| News | R01 à R05 | 6,91 | 6,77 | 6,19 |
| Animals | R06 à R10 | 5,25 | 6,13 | 5,87 |
| Movies | R11 à R15 | 5,72 | 5,13 | 5,67 |
| Health | R16 à R20 | 4,98 | 4,83 | 4,66 |
| Sports | R21 à R25 | 5,93 | 5,89 | 5,16 |
| Travel | R26 à R30 | 6,19 | 6,09 | 6,10 |

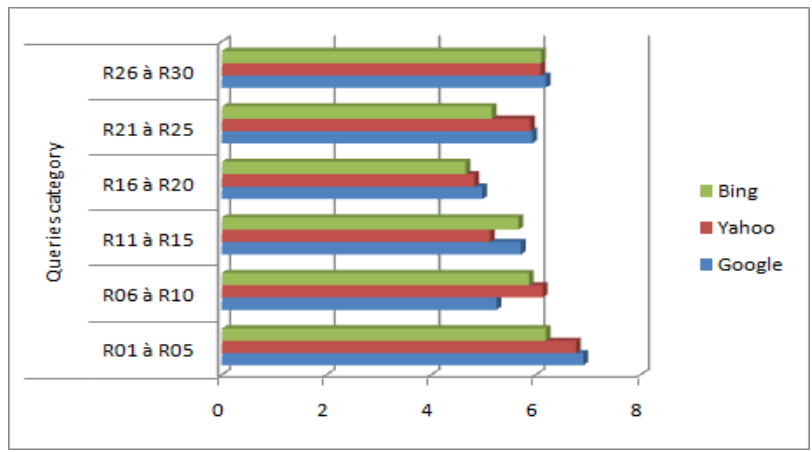

Figure 6. Evaluation of the results relevance according to the query

The analysis of the results obtained show that the Google search engine ranks first in terms of relevance of results according to the query, and that for the 5 query categories of the 6 available categories. This finding may be explained by a possible match or an unintended complicity between the formula that we proposed and the mechanism Google uses to rank results. We also note that the scores of the 'health' category are below average for the three search engines, this is due to the fact that the queries in this category contain few of words, which decreases the words for which we calculate the number of occurrence and thus weaken the final score.

## 7. CONCLUSION

In this paper we are interested in proposing a new approach based on the context for evaluating information retrieval systems. A deep investigation of the work done in the field of classic evaluation of this type of system allows us to identify limits and shortcomings encountered during the evaluation process. We have therefore defined three classes of problems, each class is related to an actor that we generally find around the evaluation process. These limits are essentially those related to the absence of the user during the evaluation, those related to the relevance judgments and finally the limits related to the corpus of documents and queries.

Our main contribution consists of the consideration of context during the evaluation at three complementary levels. First the context of the system is considered by estimating the ability of the search tool to eliminate the dead links, redundant results and parasites pages. In a second level our approach takes into account the query context based on an incremental formula for calculating the relevance of the returned results according to the sent query. The last level of the approach takes into consideration the user's judgments via his static and dynamic context. Finally, a synthesis of the three levels of contextual evaluation was proposed.

The application of the proposed approach to the search engines evaluation was used to demonstrate its applicability for real research tools. This study which is certainly far from exhaustive, nevertheless gives a snapshot of the search engines performance and the relevancy of results that they return. We note also that nothing in this study helps to explain the massive user preference for the Google search engine because, overall, Google and Yahoo have performance roughly equivalent. We must therefore assume that the reasons are criteria other than the pure relevance.

Finally, this study paves the way for diverse perspectives; the most important of them is to enlarging the application field of the realized research. It would be interesting to test the proposed approach to evaluate personalized search tools and enrich the obtained results with search engines.